\documentclass[12pt,a4paper]{article}
\usepackage{graphicx}
\usepackage{amsmath, amssymb, amsthm}
\usepackage[left=1in, right=1in, top=1in, bottom=1in]{geometry}
\usepackage[dvipsnames]{xcolor} 
\usepackage{soul}
\usepackage[braket]{qcircuit}
\usepackage{plain}
\usepackage{subcaption}

\usepackage{authblk}

\DeclareMathOperator{\Tr}{Tr}
\title{\textbf{The effect of noisy environment on Secure Quantum Teleportation of uni-modal Gaussian states}} 

\author[1]{S. Mehrabankar}
\author[2, 3]{P. Mahmoudi}
\author[4]{F. Abbasnezhad}
\author[4, 5]{D. Afshar}
\author[6]{A. Isar}
\affil[1]{Departamento de Física Teórica and IFIC, Universidad de Valencia-CSIC, 
46100 Burjassot (Valencia), Spain}
\affil[2]{Department of Surface and Plasma Science, Faculty of Mathematics and Physics, Charles University, V Holešovičkách 2, Praha 8, Czech Republic}
\affil[3] {Department of Optics, Palacký University, 17. listopadu 12, 771 46 Olomouc, Czech Republic}
\affil[4]{Department of Physics, Faculty of Science, Shahid Chamran University of Ahvaz, Ahvaz, Iran}
\affil[5]{Center for Research on Laser and Plasma, Shahid Chamran University of Ahvaz, Ahvaz, Iran}
\affil[6]{Department of Theoretical Physics, National Institute of Physics and Nuclear Engineering, Bucharest-Magurele, Romania}

\date{}

\begin{document}

	\maketitle{}
 \textbf{ABSTRACT} Quantum communication networks can be built on quantum teleportation, which is the transmission of an unknown quantum state from a sending station to a remote receiving station supported by entangled states and classical communication. We use a continuous variable two-mode squeezed vacuum state as a resource state for the quantum teleportation. This state is shared by Alice and Bob, and their system comes into contact with a squeezed thermal environment. The conditions for a secure quantum teleportation require a teleportation fidelity larger than 2/3 and two-way steering of the resource state. 
We investigate the time evolution of the steering and the fidelity of teleportation in order to determine the values of the parameters required for a successful secure quantum teleportation of a coherent Gaussian state. We show that the temperature, dissipation rate and squeezing parameter of the squeezed thermal reservoir limit the feasible duration for secure quantum teleportation, while by increasing the squeezing parameter of the initial state one can effectively expand the temporal range for a successful secure quantum teleportation. 
 
\section{Introduction}

Considerable attention has been paid to quantum teleportation in recent years, both as a fundamental quantum information protocol and as a component of quantum technologies \cite{vaidman1994teleportation, bouwmeester1997experimental, boschi1998experimental, braunstein1998teleportation, furusawa1998unconditional, bowen2003experimental, zhang2003quantum, riebe2004deterministic, yonezawa2004demonstration, sherson2006quantum, yukawa2008high, yin2012quantum, afshar2020two}. Through the utilization of conventional communication and the sharing of an entangled state as a quantum resource, quantum information can be transferred from one location to another via the process of quantum teleportation \cite{bennett1993teleporting}. The effective realisation of quantum teleportation is dependent on the shared state entanglement. A crucial component of the quantum internet is high-fidelity quantum teleportation, that enables information to be transmitted with a fidelity that exceeds the classical limit \cite{massar1995optimal, horodecki1999general, banaszek2000optimal, albeverio2002optimal, kimble2008quantum}. Quantum information needs to be securely sent over vast distances via a quantum communication network based on quantum teleportation \cite{furusawa2007quantum, pirandola2015advances}. 

Bennett \cite{bennett1993teleporting} created the first quantum teleportation protocol for discrete variable systems, that was later expanded to continuous variable systems by Vaidman, Braunstein and Kimble \cite{vaidman1994teleportation, braunstein1998teleportation, kimble2008quantum}. The Braunstein-Kimble protocol \cite{massar1995optimal, horodecki1999general} enables traditional quantum teleportation. The legal parties in continuous variable quantum teleportation can establish Gaussian entanglement via lossy channels \cite{yukawa2008high, banaszek2000optimal, kimble2008quantum}. Furthermore, the local oscillator can be used as a filter for coherent detection to reduce background noise \cite{wu2021passive, wang2019feasibility}. The transmission of information via the free space channel is currently gaining popularity due to its unique advantages. Gaussian states are crucial in quantum optics and quantum information processing and transmission since they can be easily produced and controlled. They are particularly important resources in quantum teleportation \cite{furusawa1998unconditional, bowen2003experimental, adesso2014continuous}.

Dissipative effects that result from interactions with the environment have an impact on real-world physical systems and they influence the quantum information processing \cite{abbasnezhad2017evolution, abbasnezhad2018markovian, afshar2016entanglement, mehrabankar2019quantum}. As a result, research on quantum teleportation in open quantum systems has garnered a lot of attention lately \cite{badziag2000local, verstraete2002fidelity, bandyopadhyay2002origin, kumar2003effect, jung2008greenberger, rao2008teleportation, yeo2009effects, hu2010noise, olivares2003optimized, olivares2003teleportation, pirandola2018teleportation, tserkis2018simulation}. In a previous paper \cite{afshar2020two} we investigated and compared the squeezed vacuum states and squeezed thermal states as initial resource states for secure quantum teleportation (SQT) in an open quantum system, consisting of two uncoupled harmonic oscillators interacting with a thermal environment. In the present paper we apply the Gorini-Kossakowski-Lindblad-Sudarshan master equation of the theory of open quantum systems based on completely positive dynamical semi-groups to characterise the time evolution of a system consisting of two bosonic modes interacting with a squeezed thermal reservoir. The state of this system is taken as a resource state for the quantum teleportation of a Gaussian coherent state. Two-way steering of the resource state and teleportation fidelity larger than 2/3 are required for SQT. We will show that the quantum steering and the fidelity of teleportation, and therefore SQT, strongly depend on the input state to be teleported and on the parameters characterising the resource state and environment. Our purpose is to determine the dependence of the intervals of the allowed time for SQT on these parameters, in the case of a common squeezed thermal environment.

The paper is organised as follows. In Sections 2 we introduce the necessary ingredients about SQT of a coherent state. Section 3 is devoted to the description of the effect of the noisy environment on SQT. In Section 4 there are discussed the obtained results and the final section contains the concluding remarks.

\section{Secure quantum teleportation}

Secure quantum teleportation is the process of transporting a quantum state without deviating and losing information. Once teleportation fidelity exceeds 2/3, teleportation is guaranteed. However, the interaction with the environment compromises the security of quantum teleportation. The shared entangled state of the sender and receiver as a resource is vital in quantum teleportation. Recently, it was demonstrated that SQT requires for the resource state a two-way steerability, that is a type of quantum correlation stronger than entanglement \cite{afshar2020two, grosshans2001quantum, he2015secure, he2011continuous, cuzminschi2021extractable, zubarev2019optimal, marian2006continuous, schrodinger1935discussion, einstein1935can}. These two conditions for two-mode Gaussian states are summarised in Subsections 2.1 and 2.2 as follows.

\subsection{Quantum fidelity of teleportation}

 The general expression of the quantum fidelity of two states, in terms of their density operators
$\rho_{1}$ and $\rho_{2}$, is given by
\begin{equation}
\label{eqn:1}
F(\rho_{1},\rho_{2})=[{\rm Tr}(\sqrt{\sqrt{\rho_{2}}\rho_{1}\sqrt{\rho_{2}}}){]}^{2}.
\end{equation}
Fidelity measures the transition probability between these two states and it has the following properties \cite{bennett1993teleporting}:

$1. ~ 0\leq F(\rho_{1},\rho_{2})\leq1$ and $
 F(\rho_{1},\rho_{2})=1$ if and only if $\rho_{1}=\rho_{2}$;
 
2. $F(\rho_{1},\rho_{2})=F(\rho_{2},\rho_{1})$ (symmetry);

3. $ F(\rho_{1},\rho_{2})={\rm Tr}(\rho_{1}\rho_{2})$ if either $\rho_{1}$ or
$\rho_{2}$ is a pure state;

4. $F(U\rho_{1}U^{\dagger},U\rho_{2}U^{\dagger})=F(\rho_{1},\rho_{2}),$ where
$U$ is a unitary transformation on the state space.
\newline The quantum fidelity is 0 when the two states are orthogonal:
$F(\rho_{1},\rho_{2})=0
\newline
\equiv <\rho_{1}\mid\rho_{2}>=0$.

Using Eq. (\ref{eqn:1}) for the quantum teleportation protocol, we can determine the teleportation fidelity, that describes the similarity between the input state (to be teleported) $\rho_{in}=\rho_{1}$
and the output (teleported) state $\rho_{out}=\rho_{2}$. As a result, the fidelity of teleportation, which takes values ranging from 0 to 1, evaluates the efficiency of quantum teleportation. In general, in an experimental setup the input and output states are not identical, so that the fidelity of teleportation is less than 1; in addition, the fidelity of teleportation is affected by the dissipation and decoherence phenomena that occur during the interaction with the environment.  
In the case of teleportation of unimodal Gaussian states, the expression (\ref{eqn:1}) of the teleportation fidelity is as follows \cite{badziag2000local, verstraete2002fidelity}:
\begin{equation}
\label{eqn:2}
    F(\rho_{in},\rho_{out})=\frac{\exp[-\frac{1}{2}(\overline{X_{out}}-\overline{X_{in}})^{\rm T}(\sigma_{in}+\sigma_{out})^{-1}(\overline{X_{out}}-\overline{X_{in}})]}{\sqrt{\Delta+\Theta}-\sqrt{\Theta}},
\end{equation}
where $\sigma_{in}$ and $\sigma_{out}$ are the covariance matrices of the input and the output state, respectively; $\overline{X_{in}}={\rm Tr}[\rho_{in}(X,P)^{\rm T}]$ and $\overline{X_{out}}={\rm Tr}[\rho_{out}(X,P)^{\rm T}]$
  are the average values of the quadrature operators of the input and output states, respectively. Moreover, $\Delta$ and $\Theta$ in \ref{eqn:2} are given by: 
  \[\Delta={\rm det}(\sigma_{in}+\sigma_{out})\geq1\]
  and
  \[\Theta=4{\rm det}(\sigma_{in}+\frac{\rm i}{2}J){\rm det}(\sigma_{out}+\frac{\rm i}{2}J),\]
  where $J=\left(\begin{array}{cc}
  	0 & 1\\
  	-1 & 0
  \end{array}\right)$.

 As input state to be teleported we use an unimodal Gaussian coherent state with the following covariance matrix
\begin{equation}
\label{eqn:3}
\sigma_{in}=\left(\begin{array}{cc}
	\frac{1}{2} & 0\\
	0 & \frac{1}{2}
\end{array}\right).
\end{equation}
In order to derive the output teleported state we employ the characteristic function approach. The quantum teleportation is accomplished by using a classical communication channel and an entangled state shared by two parties, Alice A and Bob B. At the initial moment of the teleportation, parties A and B are assumed to be in an entangled squeezed vacuum state, described by the following characteristic function \cite{he2015secure, he2011continuous, zubarev2019optimal, cuzminschi2021quantum}:
\begin{equation}
\label{eqn:4}
\chi_{AB}(\lambda_{A},\lambda_{B})=\exp\{-\frac{1}{2}\Lambda^{\rm T}\sigma(0)\Lambda-{\rm i}\Lambda^{\rm T}\overline{X_{AB}(0)}\},
\end{equation}
where $\sigma(0)$ is the initial covariance matrix of the bipartite system AB, $\lambda_{A}=-\frac{\rm i}{\sqrt{2}}(x_{A}+{\rm i}p_{A})$, $\lambda_{B}=-\frac{\rm \rm i}{\sqrt{2}}(x_{B}+{\rm i}p_{B})$ and $\Lambda=-\frac{\rm i}{\sqrt{2}}(x_{A},x_{B},p_{A},p_{B})^{\rm T}$; $\overline{X_{AB}(0)}$ defined by:
\begin{equation}
\label{eqn:5}
\overline{X_{AB}(0)}={\Tr}[\rho(0)(X_{A},P_{A},X_{B},P_{B})^{\rm T}]
\end{equation}
 is the average value of the quadrature position and momentum operators ($X_{j}=\frac{1}{\sqrt{2}}(a_{j}+a_{j}^{\dagger}$) and, respectively, $P_{j}=\frac{-\rm i}{\sqrt{2}}(a_{j}-a_{j}^{\dagger}$), $j=A,B$), where $a_{j}$, $a_{j}^{\dagger}$ are the annihilation and creation operators, and $\rho(0)$ is the initial density operator. 
If the evolution in time of the bipartite system is Gaussian, then the shared entangled state remains Gaussian during the interaction with the surrounding environment and it has the following characteristic function at the moment of time $t$:
\begin{equation}
\label{eqn:6}
\chi_{AB}^{t}(\lambda_{A},\lambda_{B})=\exp\{-\frac{1}{2}\Lambda^{\rm T}\sigma(t)\Lambda-{\rm i}\Lambda^{\rm T}\overline{{X}_{AB}(t)}\},
\end{equation} where 
\begin{equation}
\label{eqn:7}
\overline{X_{AB}(t)}={\rm Tr}[\rho(t)(X_{A},P_{A},X_{B},P_{B})^{\rm T}]\end{equation} is the time-dependent averaged value of the quadrature operators. 
We denote the covariance matrix of the shared state by \cite{zubarev2019optimal, cuzminschi2021quantum}:
\begin{equation}
\label{eqn:8}
\sigma(t)=\left(\begin{array}{cc}
	A(t) & C(t)\\
	C^{\rm T}(t) & B(t)
\end{array}\right),
\end{equation}
where $ A(t)=\left(\begin{array}{cc}
	A_{11} & A_{12}\\
	A_{12} & A_{22}
\end{array}\right)$, $B(t)=\left(\begin{array}{cc}
B_{11} & B_{12}\\
B_{12} & B_{22}
\end{array}\right)$, and $C(t)=\left(\begin{array}{cc}
C_{11} & C_{12}\\
C_{21} & C_{22}
\end{array}\right)$.

The expression of the covariance matrix of the output state can be derived by employing the procedure used in Ref. \cite{zubarev2019optimal} and it has the following form:
\begin{equation}
\label{eqn:9}
\sigma_{out}=\left(\begin{array}{cc}
	A_{11}+B_{11}-2C_{11}+\frac{1}{2} & A_{12}-B_{12}+C_{12}-C_{21}\\
	A_{12}-B_{12}+C_{12}-C_{21} & A_{22}+B_{22}+2C_{22}+\frac{1}{2}
\end{array}\right).
\end{equation}
For the sake of simplicity, we will assume that $\overline{X_{out}}=\overline{X_{in}}$. Then the fidelity of teleportation (\ref{eqn:2}) reduces to
\begin{equation}
\label{eqn:10}
F(\rho_{in},\rho_{out})=\frac{1}{\sqrt{\triangle}},
\end{equation}
with
\[\triangle=\det(\sigma_{in}+\sigma_{out})=1+X+Y+XY-Z^{2}\]
and $\Theta=0$,
where $X$, $Y$ and $Z$ are given by:
\begin{equation}
\label{eqn:11}
	\begin{split}
X=A_{11}+B_{11}-2C_{11},\\
Y=A_{22}+B_{22}+2C_{22},\\
Z=A_{12}-B_{12}+C_{12}-C_{21}.
\end{split}
\end{equation}
By denoting
\begin{equation}
\label{eq:12}
\Sigma=\left(\begin{array}{cc}
	X & Z\\
	Z & Y
\end{array}\right),
\end{equation}
the covariance matrix of the output state in Eq. (\ref{eqn:9}) can be written as:
\begin{equation}
\label{eqn:13}
\sigma_{out}=\sigma_{in}+\Sigma
\end{equation}
and $\triangle$ is as follows:
\begin{equation}
\label{eqn:14}
\triangle=1+{\rm Tr}~\Sigma+\det\Sigma.
\end{equation}

\subsection{Two-way quantum steering}

Steering is a kind of quantum correlation introduced by Schr\"odinger in the context of the EPR paradox \cite{marian2006continuous, schrodinger1935discussion}. Quantum steering refers to the ability of one party in a bipartite system to change the state of the other party by using local measurements. Since the effects of local measurements on the two subsystems are, in general, different, hence, the steering is fundamentally asymmetric \cite{einstein1935can}. Due to this property, steerable states are useful in quantum processes where the result of the measurement on one party is not trustable, like quantum key distribution \cite{wiseman2007steering, branciard2012one}. Steerable states are also useful for channel discrimination and SQT \cite{quintino2015inequivalence, jones2007entanglement}. 

We only consider two-mode squeezed vacuum states in this paper, which are a subset of continuous variable states defined by a Gaussian Wigner function. Besides, they are described in phase space by a $4\times4$ covariance matrix, as shown in Eq. (\ref{eqn:8}). It has been demonstrated that a bipartite Gaussian state is steerable if Alice can generate different states for the Bob subsystem, by using different local measurements on her subsystem, and if the local hidden state model cannot generate these states \cite{quintino2015inequivalence, jones2007entanglement, piani2015necessary, giedke2002characterization}. According to this model, a bipartite Gaussian state with the covariance matrix $\sigma_{AB}$ is  $A\rightarrow B$ steerable in terms of Alice measurements, if and only if the following inequality is not satisfied \cite{adesso2014continuous}:
\begin{equation}
\label{eqn:15}
\sigma_{AB}+\frac{\rm i}{2}(0_{A}\oplus J)\geq 0, 
\end{equation}
which is equivalent to the following conditions:
\begin{equation}
\label{eqn:16}
A>0,~~ M_{\sigma}^{B}+\frac{\rm i}{2}J\geq 0, 
\end{equation}
where $M_{\sigma}^{B}=B-C^{\rm T}A^{-1}C$ is the Schur complement of $A$ in the covariance matrix $\sigma_{AB}$ \cite{quintino2015inequivalence, jones2007entanglement}. Since the covariance matrix of each subsystem has to be physical, the first condition in Eqs. (\ref{eqn:16}) is always satisfied. As a result, $\sigma_{AB}$ is $A\rightarrow B$ steerable only if the second condition is not satisfied \cite{marian2006continuous, piani2015necessary}. This condition can be written as \cite{giedke2002characterization, fiuravsek2007gaussian}:
\begin{equation}
\label{eqn:17}
\nu^{B}\geq\frac{1}{2},
\end{equation}
where $\nu^{B}$ is the symplectic eigenvalue of $M_{\sigma}^{B}$. Furthermore, the measure of  $A\rightarrow B$ steerability can be determined by quantifying the degree of violation in Eqs. (\ref{eqn:16}) as follows:
\begin{equation}
\label{eqn:18}
S^{A\rightarrow B}(\sigma_{AB})={\rm max}\{0,-\ln(2\nu^{B})\}.
\end{equation} 
This quantity is invariant under symplectic transformations and vanishes if and only if  $\sigma_{AB}$ is non-steerable. Likewise, the $B\rightarrow A$ steerability can be obtained by swapping the roles of $A$ and $B$ and replacing the symplectic eigenvalue of the Schur complement of $B$ by that one of $A$ in Eq. (\ref{eqn:18}). The measure of steering (\ref{eqn:18}) has a simple analytical expression in the case of two-mode Gaussian states \cite{williamson1936algebraic}:
\begin{equation}
\label{eqn:19}
S^{A\rightarrow B}(\sigma_{AB})={\rm max}\{0,\frac{1}{2}\ln\frac{\det A}{4\det\sigma_{AB}}\}.
\end{equation}
As a result, the necessary condition for SQT of coherent states is \cite{kogias2015quantification}:
\[\mathcal{L}>0,\]
where 
\begin{equation}
\label{eqn:20}
\mathcal{L}={\rm min}\{S^{A\rightarrow B},S^{B\rightarrow A},F-\frac{2}{3}\}.
\end{equation}

\section{The effect of noisy environment on SQT}

We consider an open quantum system composed of two uncoupled bosonic modes in a squeezed bosonic environment. The Hamiltonian that describes the system is as follows \cite{wallsquantum}:
\begin{equation}
\label{eqn:21}
H=\sum_{i=1}^{2}\hbar\omega a_{i}^{\dagger}a_{i}+\sum_{j=1}^{N}\hbar\omega b_{j}^{\dagger}b_{j}+\sum_{i=1}^{2}\hbar(G a_{i}^{\dagger}+ G^{\dagger}a_{i}),\\    G=\sum_{j=1}^{2}g_{j}b_{j},
\end{equation}
where $a_{i}^{\dagger}$  
 and $a_{i}$, and $b_{j}^{\dagger}$ and $b_{j}$ are, respectively, the creation and annihilation operators of the system and of the environment, and $g_{j}$ are the strength of the interaction between system and environment. 
The environment is typically composed of radiation field modes that can be provided in various states such as vacuum, squeezed, thermal, and squeezed thermal. Therefore, the environment is characterized by the correlation functions between environment operators at different times as follows \cite{wallsquantum}:
\begin{equation}
\label{eqn:22}
	\begin{split}
<b_{j}^{\dagger}(t)b_{j}(t^{\prime})>=N_{j}\delta(t-t^{\prime}),~
<b_{j}(t)b_{j}^{\dagger}(t^{\prime})>=(N_{j}+1)\delta(t-t^{\prime}),\\
<b_{j}(t)b_{j}(t^{\prime})>=M_{j}\delta(t-t^{\prime}),~<b_{j}^{\dagger}(t)b_{j}^{\dagger}(t^{\prime})>=M_{j}^{\ast}\delta(t-t^{\prime}),
\end{split}
\end{equation}
where $N_{j}$ and  $M_{j}$ are the mean photon numbers and the squeezing parameters of the environment, respectively \cite{he2015secure}. For $M_{j}=N_{j}=0$, the environment is in a vacuum state, for  $N_{j}\neq0, M_{j}=0$, it is in a thermal state, and for $M_{j}\neq0, N_{j}\neq0$, it is in a squeezed thermal state. The Heisenberg uncertainty relation imposes the following constraint on $N_{j}$ and $M_{j}$:
\[\mid M_{j}\mid^{2}\leq N_{j}(N_{j}+1).\]

In the case of a common squeezed thermal environment (correlated noisy channel) one has
\[N_{1}=N_{2}=N, ~M_{1}=M_{2}=M\]
and the irreversible time evolution of the considered open system is described, in the Markovian approximation, by the following Lindblad master equation in the interaction representation \cite{xiang2015nonclassical}:

\begin{eqnarray}
\label{eqn:23}
	\frac{\partial}{\partial t}\rho=\sum_{i=1}^{2}\frac{\gamma}{2}\{N(2a_{i}^{\dagger}\rho a_{i}-a_{i}a_{i}^{\dagger}\rho-\rho a_{i}a_{i}^{\dagger})+(N+1)(2a_{i}\rho a_{i}^{\dagger}-a_{i}^{\dagger}a_{i}\rho-\rho a_{i}^{\dagger}a_{i})\}\nonumber\\
		-\sum_{i\neq j=1}^{2}\frac{\gamma}{2}\{
			M^{\ast}(2a_{i}\rho a_{j}-a_{i}a_{j}\rho
		-\rho a_{i}a_{j})
		+M(2a_{i}^{\dagger}\rho a_{j}^{\dagger}-a_{i}^{\dagger}a_{j}^{\dagger}\rho-\rho a_{i}^{\dagger}a_{j}^{\dagger})\},
\end{eqnarray}
where $\gamma$ is the environment field decay rate. 

The asymptotic covariance matrix of the two bosonic modes is defined by the bath parameters only, and it is given by \cite{kogias2015quantification}:
\begin{equation}
\label{eqn:25}
\sigma(\infty)=\left(\begin{array}{cc}
	(N+1/2)I_{2} & M\sigma_{z}\\
	M\sigma_{z} & (N+1/2)I_{2}
\end{array}\right),~ \sigma_{z}=\left(\begin{array}{cc}
	1 & 0\\
	0 & -1
\end{array}\right).
\end{equation}
Here 
\begin{eqnarray}
\label{eqn:26}
N=n_{th}(\cosh^2 R+\sinh^2 R)+\sinh^2 R, \nonumber\\
M= -(2n_{th}+1)\cosh R\sinh R
\end{eqnarray}
and $R$ is the squeezing parameter of the reservoir. 
The average thermal photon number of the bath is given by (we set $\hbar=1$, $\omega=1$ and the Boltzmann constant $k_{B}=1$) 
\begin{equation}
\label{eqn.27}
n_{th}=\frac{1}{2}\left(\operatorname{coth} \frac{1}{2 T}-1\right),
\end{equation}
where $T$ is the temperature of the reservoir.

The solution of Eq. (\ref{eqn:23}) in terms of the covariance matrix is given by \cite{afshar2020two, xiang2015nonclassical}:
\begin{equation}
\label{eqn:31}
\sigma(t)=\Gamma\sigma(0)+(I_{4}-\Gamma)\sigma(\infty),
\end{equation}
where $\ensuremath{\Gamma=e^{-\gamma t}I_{4}}$;
$\sigma(0)$ and $\sigma(\infty)$ are the covariance matrices of the initial and final states, respectively. 
We take a squeezed vacuum state as initial state with the following covariance matrix \cite{kogias2015quantification}:
\begin{equation}
\label{eqn:32}
\sigma(0)=\frac{1}{2}\left(\begin{array}{cccc}
	\cosh 2r & 0 & \sinh 2r & 0\\
	0 & \cosh 2r & 0 & -\sinh 2r\\
	\sinh 2r & 0 & \cosh 2r & 0\\
	0 & -\sinh 2r & 0 & \cosh 2r
\end{array}\right),
\end{equation}
where $r$ is the squeezing parameter.

\section{Results and discussions}

In this study, we focus on utilizing a two-mode squeezed vacuum state as a resource state for quantum teleportation. This state is shared between Alice and Bob, while their system interacts with a common squeezed thermal environment. Our primary objective is to investigate how the fidelity of the teleportation process and quantum steering evolve in time, aiming to identify the necessary parameters for a successful SQT of an one-mode Gaussian coherent state.
We recall that to ensure a SQT two conditions have to be met, namely the teleportation fidelity needs to be larger than 2/3 and the resource state has to be two-way steerable. 

\begin{figure}[htbp]
  \centering
  \begin{subfigure}{0.45\textwidth}
    \centering
    \includegraphics[width=\textwidth]{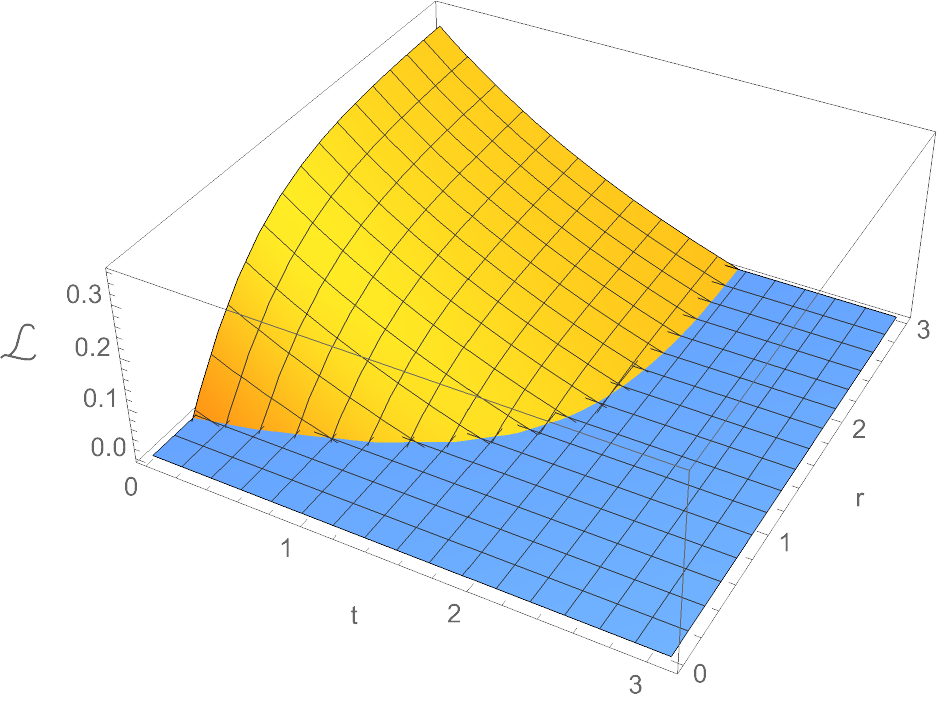}
    \caption{}
    \label{1(a)}
  \end{subfigure}
  \hfill
  \begin{subfigure}{0.45\textwidth}
    \centering
    \includegraphics[width=\textwidth]{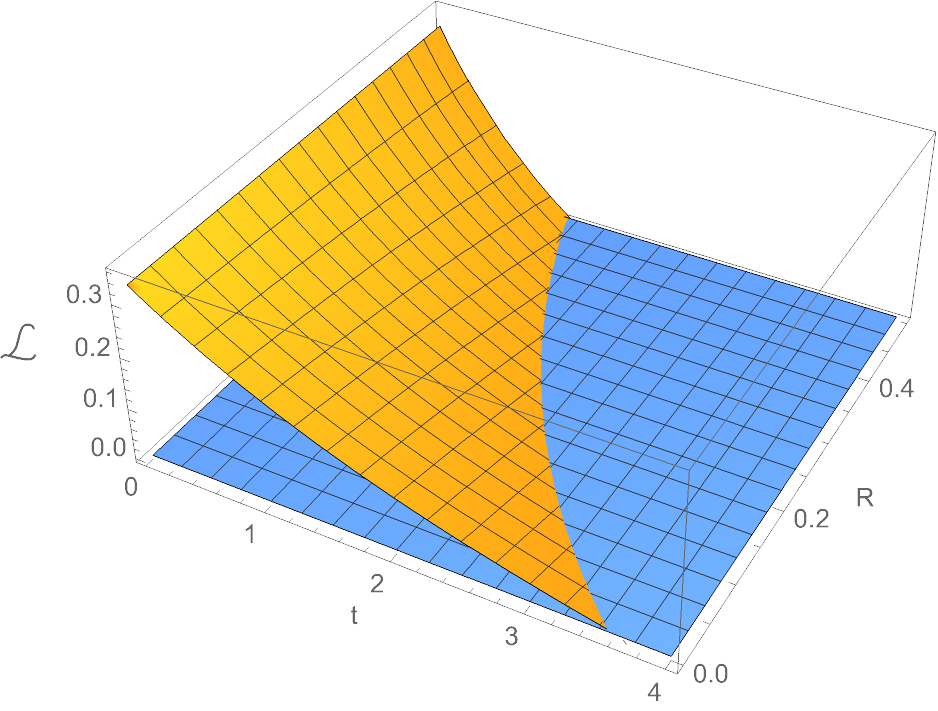}
    \caption{}
    \label{1(b)}
  \end{subfigure}
  \caption{The regions where $\mathcal{L}>0$ versus time $t$ and: (a) squeezing $r$ of the initial state for $R=0.1, T=1$; (b) squeezing $R$ of the squeezed thermal environment for $r=3, T=0.7$.  We set $\gamma=0.1.$}
  \label{Fig1}
\end{figure}

In Fig.\ref{1(a)} we present a plot illustrating the relationship between the allowed time for SQT and the squeezing parameter of the initial resource state. We observe that as the squeezing parameter increases, the range of admissible time for SQT expands significantly. This indicates that by enhancing the squeezing parameter, one can effectively extend the duration during which a successful SQT is achievable. At the same time we notice that for relatively small values of the squeezing parameter the SQT cannot be realised.

Fig.\ref{1(b)} depicts a plot that illustrates how the conditions for a SQT depend on time and the squeezing parameter of the squeezed thermal reservoir. Compared to the previous case, we observe that as the squeezing parameter of the environment increases, the temporal window for a successful SQT reduces. In addition, as it will be shown in the following, for relatively large values of the squeezing parameter of the thermal reservoir, the SQT cannot be realised.

\begin{figure}[htbp]
  \centering
  \begin{subfigure}{0.45\textwidth}
    \centering
    \includegraphics[width=\textwidth]{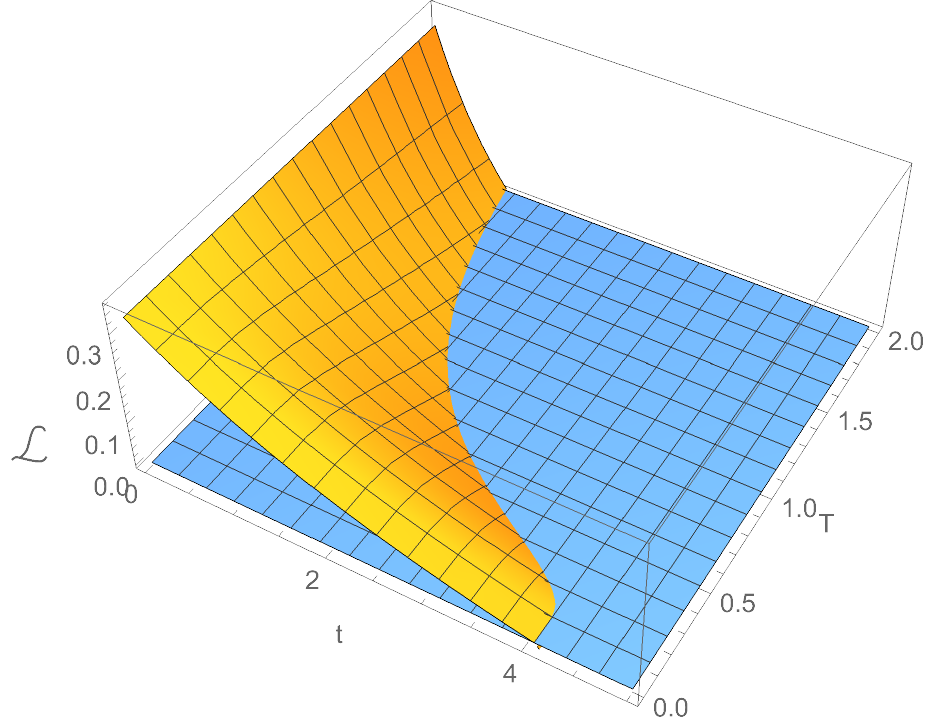}
    \caption{}
    \label{2a}
  \end{subfigure}
  \hfill
  \begin{subfigure}{0.45\textwidth}
    \centering
    \includegraphics[width=\textwidth]{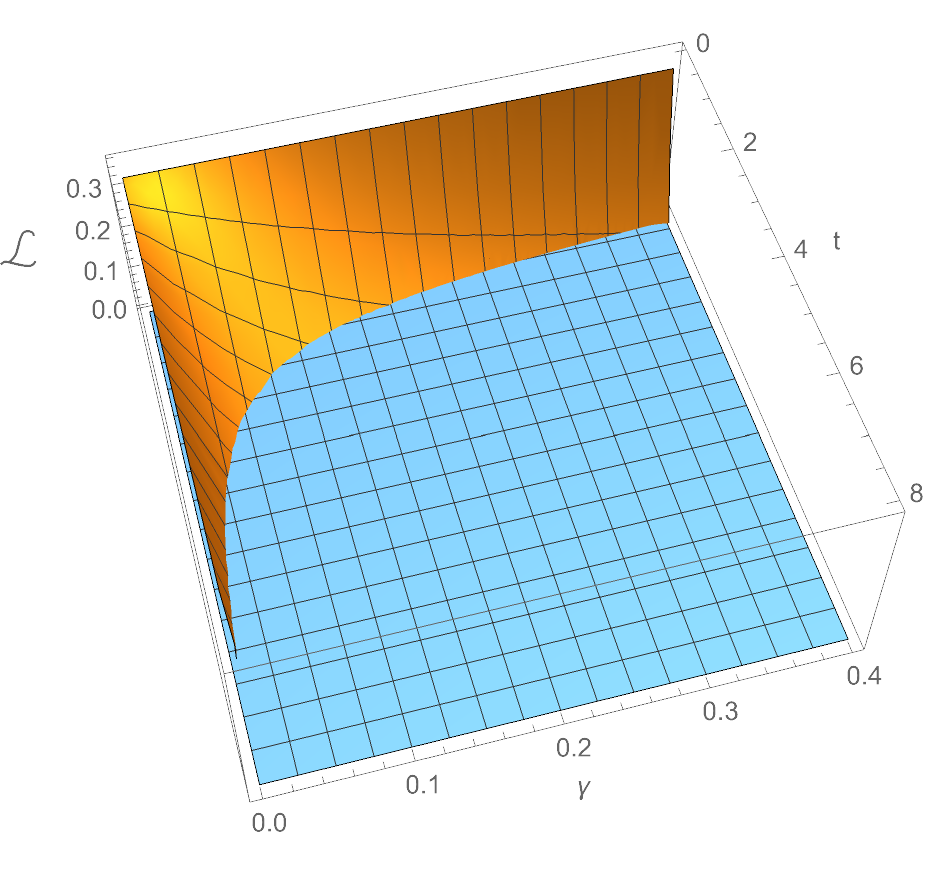}
    \caption{}
    \label{2b}
  \end{subfigure}
  \caption{The regions where $\mathcal{L}>0$ versus time $t$ and: (a) temperature $T$ of the squeezed thermal environment for $R=0.2, \gamma=0.1$; (b) dissipation rate $\gamma$ for $T=1.1, R=0.09$.  We set $r=3.$}
  \label{Fig2}
\end{figure}

In Fig. \ref{Fig2} we present the plots that depict the regions for which the conditions for a SQT is fulfilled as a function of time and the temperature of the squeezed thermal reservoir \ref{2a}  and the dissipation rate \ref{2b}.  As expected, we observe that if both the temperature of the reservoir and the dissipation rate increase, then the duration within which SQT can be realised decreases. 

 \begin{figure}[htbp]
 	\centering
 	\includegraphics[scale=0.5]{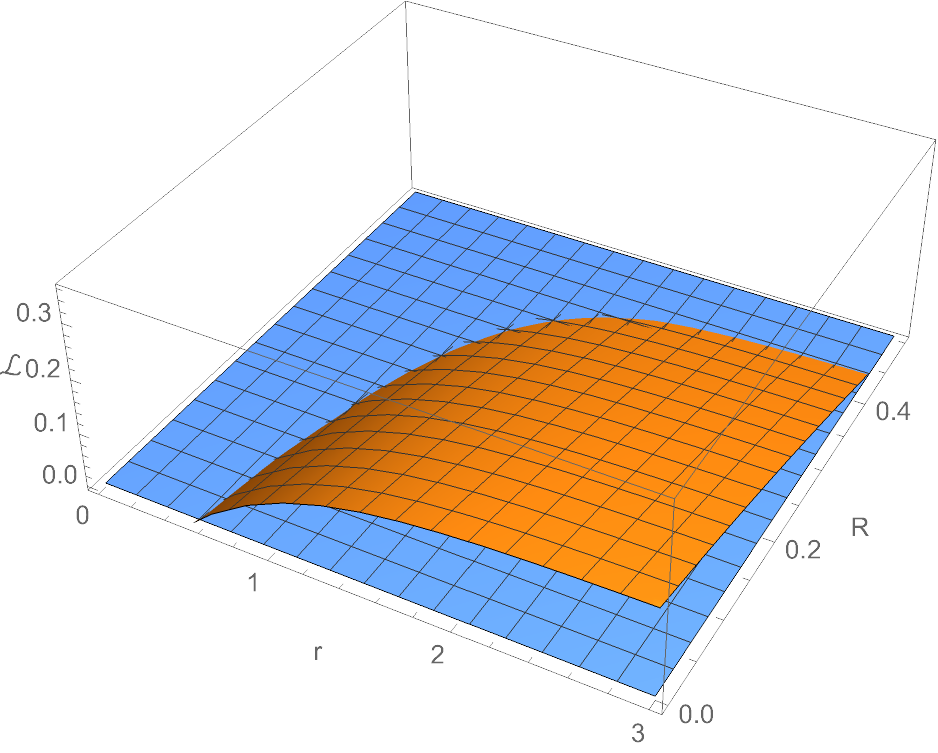}
 	\caption{The regions where $\mathcal{L}>0$ versus squeezing $r$ of the initial state and squeezing $R$ of the squeezed thermal environment at the moment of time $t=1$ for $T=1, \gamma=0.1.$}
  \label{Fig3}
 \end{figure}

In Fig. \ref{Fig3} we depict the region where the conditions for a SQT are fulfilled as a function of the squeezing parameters of the initial state and the environment, at a given moment of time. We observe again that the increasing of the squeezing parameter of the initial state supports the fulfilment of the conditions of SQT, while the increasing of the squeezing parameter of the environment is detrimental to the fulfilment of the conditions of SQT. 
In addition, we notice that for relatively small values of the squeezing parameter of the initial state and for relatively large values of the squeezing parameter of the thermal reservoir, SQT cannot be realised. 
 
 \section{Conclusion}
 
The present study has provided valuable insights into the temporal dynamics and the influence of key parameters on SQT. Through our investigations, we have observed significant correlations and trends that shed light on the feasibility and limitations of this quantum communication protocol.

One of our main findings is the description of the influence of environmental factors on the allowed time for SQT. Specifically, we have observed that the nature of the squeezed thermal environment plays a crucial role in determining the temporal window for a successful secure teleportation. 
Namely, we examined the impact of the temperature, dissipation rate and squeezing parameter of the squeezed thermal reservoir on the allowed time for SQT and we noticed that all these parameters characterising the environment limit the feasible duration for SQT.
We have also investigated the impact of the squeezing parameter of the initial state on the allowed time for SQT and the obtained results indicate that by increasing the squeezing of the initial state can effectively expand the temporal range for a successful SQT. 

These findings underscore the significance of carefully considering environmental conditions and parameters when designing and implementing SQT protocols. By optimizing the squeezing parameters and mitigating diffusion and dissipation effects, researchers can enhance the reliability and robustness of quantum teleportation, enabling more extended periods of secure information transfer.

In summary, our study contributes to the understanding of temporal constraints and parameter dependencies for SQT. These findings could have implications for the development of practical quantum communication systems and provide guidance for future research aiming to improve the efficiency and security of quantum teleportation protocols. 

\section*{Acknowledgments}
The author S. Mehrabankar would like to acknowledge the support by the Spanish MCIN/AEI/10.13039/501100011033 grant PID2020-113334GB-I00, Generalitat Valenciana grant CIPROM/2022/66, the Ministry of Economic Affairs and Digital Transformation of the Spanish Government through the QUANTUM ENIA project call - QUANTUM SPAIN project, and by the European Union through the Recovery, Transformation and Resilience Plan - NextGenerationEU within the framework of the Digital Spain 2026 Agenda, and by the CSIC Interdisciplinary Thematic Platform (PTI+) on Quantum Technologies (PTI-QTEP+). This project has also received funding from the European Union’s Horizon 2020 research and innovation program under grant agreement 101086123-CaLIGOLA. A. Isar acknowledges the support by the Romanian Ministry of Research, Innovation and Digitization, through the Project PN 23 21 01 01/2023. D.Afshar would like to thank Shahid Chamran University of Ahvaz for the grant no. SCU.SP1402.812.



\begin{thebibliography}{10}

\bibitem{vaidman1994teleportation}
Lev Vaidman.
\newblock Teleportation of quantum states.
\newblock {\em Physical Review A}, 49(2):1473, 1994.

\bibitem{bouwmeester1997experimental}
Dik Bouwmeester, Jian-Wei Pan, Klaus Mattle, Manfred Eibl, Harald Weinfurter,
  and Anton Zeilinger.
\newblock Experimental quantum teleportation.
\newblock {\em Nature}, 390(6660):575--579, 1997.

\bibitem{boschi1998experimental}
Danilo Boschi, Salvatore Branca, Francesco De~Martini, Lucien Hardy, and Sandu
  Popescu.
\newblock Experimental realization of teleporting an unknown pure quantum state
  via dual classical and einstein-podolsky-rosen channels.
\newblock {\em Physical Review Letters}, 80(6):1121, 1998.

\bibitem{braunstein1998teleportation}
Samuel~L Braunstein and H~Jeff Kimble.
\newblock Teleportation of continuous quantum variables.
\newblock {\em Physical Review Letters}, 80(4):869, 1998.

\bibitem{furusawa1998unconditional}
Akira Furusawa, Jens~Lykke S{\o}rensen, Samuel~L Braunstein, Christopher~A
  Fuchs, H~Jeff Kimble, and Eugene~S Polzik.
\newblock Unconditional quantum teleportation.
\newblock {\em science}, 282(5389):706--709, 1998.

\bibitem{bowen2003experimental}
Warwick~P Bowen, Nicolas Treps, Ben~C Buchler, Roman Schnabel, Timothy~C Ralph,
  Hans-A Bachor, Thomas Symul, and Ping~Koy Lam.
\newblock Experimental investigation of continuous-variable quantum
  teleportation.
\newblock {\em Physical Review A}, 67(3):032302, 2003.

\bibitem{zhang2003quantum}
Tian~Cai Zhang, KW~Goh, CW~Chou, P~Lodahl, and H~Jeff Kimble.
\newblock Quantum teleportation of light beams.
\newblock {\em Physical Review A}, 67(3):033802, 2003.

\bibitem{riebe2004deterministic}
Mark Riebe, H~H{\"a}ffner, CF~Roos, W~H{\"a}nsel, J~Benhelm, GPT Lancaster,
  TW~K{\"o}rber, C~Becher, Ferdinand Schmidt-Kaler, DFV James, et~al.
\newblock Deterministic quantum teleportation with atoms.
\newblock {\em Nature}, 429(6993):734--737, 2004.

\bibitem{yonezawa2004demonstration}
Hidehiro Yonezawa, Takao Aoki, and Akira Furusawa.
\newblock Demonstration of a quantum teleportation network for continuous
  variables.
\newblock {\em Nature}, 431(7007):430--433, 2004.

\bibitem{sherson2006quantum}
Jacob~F Sherson, Hanna Krauter, Rasmus~K Olsson, Brian Julsgaard, Klemens
  Hammerer, Ignacio Cirac, and Eugene~S Polzik.
\newblock Quantum teleportation between light and matter.
\newblock {\em Nature}, 443(7111):557--560, 2006.

\bibitem{yukawa2008high}
Mitsuyoshi Yukawa, Hugo Benichi, and Akira Furusawa.
\newblock High-fidelity continuous-variable quantum teleportation toward
  multistep quantum operations.
\newblock {\em Physical Review A}, 77(2):022314, 2008.

\bibitem{yin2012quantum}
Juan Yin, Ji-Gang Ren, He~Lu, Yuan Cao, Hai-Lin Yong, Yu-Ping Wu, Chang Liu,
  Sheng-Kai Liao, Fei Zhou, Yan Jiang, et~al.
\newblock Quantum teleportation and entanglement distribution over
  100-kilometre free-space channels.
\newblock {\em Nature}, 488(7410):185--188, 2012.

\bibitem{afshar2020two}
Davood Afshar, Farkhondeh Abbasnezhad, Somayeh Mehrabankar, and Aurelian Isar.
\newblock Two-mode gaussian states as resource of secure quantum teleportation
  in open systems.
\newblock {\em Chinese Journal of Physics}, 68:419--425, 2020.

\bibitem{bennett1993teleporting}
Charles~H Bennett, Gilles Brassard, Claude Cr{\'e}peau, Richard Jozsa, Asher
  Peres, and William~K Wootters.
\newblock Teleporting an unknown quantum state via dual classical and
  einstein-podolsky-rosen channels.
\newblock {\em Physical review letters}, 70(13):1895, 1993.

\bibitem{massar1995optimal}
Serge Massar and Sandu Popescu.
\newblock Optimal extraction of information from finite quantum ensembles.
\newblock {\em Physical review letters}, 74(8):1259, 1995.

\bibitem{horodecki1999general}
Micha{\l} Horodecki, Pawe{\l} Horodecki, and Ryszard Horodecki.
\newblock General teleportation channel, singlet fraction, and
  quasidistillation.
\newblock {\em Physical Review A}, 60(3):1888, 1999.

\bibitem{banaszek2000optimal}
Konrad Banaszek.
\newblock Optimal quantum teleportation with an arbitrary pure state.
\newblock {\em Physical Review A}, 62(2):024301, 2000.

\bibitem{albeverio2002optimal}
Sergio Albeverio, Shao-Ming Fei, and Wen-Li Yang.
\newblock Optimal teleportation based on bell measurements.
\newblock {\em Physical Review A}, 66(1):012301, 2002.

\bibitem{kimble2008quantum}
H~Jeff Kimble.
\newblock The quantum internet.
\newblock {\em Nature}, 453(7198):1023--1030, 2008.

\bibitem{furusawa2007quantum}
Akira Furusawa and Nobuyuki Takei.
\newblock Quantum teleportation for continuous variables and related quantum
  information processing.
\newblock {\em Physics reports}, 443(3):97--119, 2007.

\bibitem{pirandola2015advances}
Stefano Pirandola, Jens Eisert, Christian Weedbrook, Akira Furusawa, and
  Samuel~L Braunstein.
\newblock Advances in quantum teleportation.
\newblock {\em Nature photonics}, 9(10):641--652, 2015.

\bibitem{wu2021passive}
Xiaodong Wu, Yijun Wang, Ying Guo, Hai Zhong, and Duan Huang.
\newblock Passive continuous-variable quantum key distribution using a locally
  generated local oscillator.
\newblock {\em Physical Review A}, 103(3):032604, 2021.

\bibitem{wang2019feasibility}
Shiyu Wang, Peng Huang, Tao Wang, and Guihua Zeng.
\newblock Feasibility of all-day quantum communication with coherent detection.
\newblock {\em Physical Review Applied}, 12(2):024041, 2019.

\bibitem{adesso2014continuous}
Gerardo Adesso, Sammy Ragy, and Antony~R Lee.
\newblock Continuous variable quantum information: Gaussian states and beyond.
\newblock {\em Open Systems \& Information Dynamics}, 21(01n02):1440001, 2014.

\bibitem{abbasnezhad2017evolution}
Farkhondeh Abbasnezhad, Somayeh Mehrabankar, Davood Afshar, and Mojtaba
  Jafarpour.
\newblock Evolution of quantum correlations in the open quantum systems
  consisting of two coupled oscillators.
\newblock {\em Quantum Information Processing}, 16:1--17, 2017.

\bibitem{abbasnezhad2018markovian}
Farkhondeh Abbasnezhad, Somayeh Mehrabankar, Davood Afshar, and Mojtaba
  Jafarpour.
\newblock Markovian thermal evolution of entanglement and decoherence of ghz
  state.
\newblock {\em The European Physical Journal Plus}, 133:1--11, 2018.

\bibitem{afshar2016entanglement}
Davood Afshar, Somayeh Mehrabankar, and Farkhondeh Abbasnezhad.
\newblock Entanglement evolution in the open quantum systems consisting of
  asymmetric oscillators.
\newblock {\em The European Physical Journal D}, 70:1--8, 2016.

\bibitem{mehrabankar2019quantum}
Somayeh Mehrabankar, Davood Afshar, and Mojtaba Jafarpour.
\newblock Quantum fidelity evolution of penning trap coherent states in an
  asymmetric open quantum system.
\newblock {\em Quantum Information \& Computation}, 19(5-6):413--423, 2019.

\bibitem{badziag2000local}
Piotr Badziag, Micha{\l} Horodecki, Pawe{\l} Horodecki, and Ryszard Horodecki.
\newblock Local environment can enhance fidelity of quantum teleportation.
\newblock {\em Physical Review A}, 62(1):012311, 2000.

\bibitem{verstraete2002fidelity}
Frank Verstraete and Henri Verschelde.
\newblock Fidelity of mixed states of two qubits.
\newblock {\em Physical Review A}, 66(2):022307, 2002.

\bibitem{bandyopadhyay2002origin}
Somshubhro Bandyopadhyay.
\newblock Origin of noisy states whose teleportation fidelity can be enhanced
  through dissipation.
\newblock {\em Physical Review A}, 65(2):022302, 2002.

\bibitem{kumar2003effect}
Deepak Kumar and PN~Pandey.
\newblock Effect of noise on quantum teleportation.
\newblock {\em Physical Review A}, 68(1):012317, 2003.

\bibitem{jung2008greenberger}
Eylee Jung, Mi-Ra Hwang, You~Hwan Ju, Min-Soo Kim, Sahng-Kyoon Yoo, Hungsoo
  Kim, DaeKil Park, Jin-Woo Son, S~Tamaryan, and Seong-Keuck Cha.
\newblock Greenberger-horne-zeilinger versus w states: Quantum teleportation
  through noisy channels.
\newblock {\em Physical Review A}, 78(1):012312, 2008.

\bibitem{rao2008teleportation}
DD~Bhaktavatsala Rao, PK~Panigrahi, and Chiranjib Mitra.
\newblock Teleportation in the presence of common bath decoherence at the
  transmitting station.
\newblock {\em Physical Review A}, 78(2):022336, 2008.

\bibitem{yeo2009effects}
Ye~Yeo, Zhe-Wei Kho, and Lixian Wang.
\newblock Effects of pauli channels and noisy quantum operations on standard
  teleportation.
\newblock {\em Europhysics Letters}, 86(4):40009, 2009.

\bibitem{hu2010noise}
Xueyuan Hu, Ying Gu, Qihuang Gong, and Guangcan Guo.
\newblock Noise effect on fidelity of two-qubit teleportation.
\newblock {\em Physical Review A}, 81(5):054302, 2010.

\bibitem{olivares2003optimized}
Stefano Olivares, Matteo~GA Paris, and Andrea~R Rossi.
\newblock Optimized teleportation in gaussian noisy channels.
\newblock {\em Physics Letters A}, 319(1-2):32--43, 2003.

\bibitem{olivares2003teleportation}
Stefano Olivares, Matteo~GA Paris, and Rodolfo Bonifacio.
\newblock Teleportation improvement by inconclusive photon subtraction.
\newblock {\em Physical Review A}, 67(3):032314, 2003.

\bibitem{pirandola2018teleportation}
Stefano Pirandola, Riccardo Laurenza, and Samuel~L Braunstein.
\newblock Teleportation simulation of bosonic gaussian channels: strong and
  uniform convergence.
\newblock {\em The European Physical Journal D}, 72:1--20, 2018.

\bibitem{tserkis2018simulation}
Spyros Tserkis, Josephine Dias, and Timothy~C Ralph.
\newblock Simulation of gaussian channels via teleportation and error
  correction of gaussian states.
\newblock {\em Physical Review A}, 98(5):052335, 2018.

\bibitem{grosshans2001quantum}
Fr{\'e}d{\'e}ric Grosshans and Philippe Grangier.
\newblock Quantum cloning and teleportation criteria for continuous quantum
  variables.
\newblock {\em Physical Review A}, 64(1):010301, 2001.

\bibitem{he2015secure}
Qiongyi He, Laura Rosales-Z{\'a}rate, Gerardo Adesso, and Margaret~D Reid.
\newblock Secure continuous variable teleportation and einstein-podolsky-rosen
  steering.
\newblock {\em Physical Review Letters}, 115(18):180502, 2015.

\bibitem{he2011continuous}
Guangqiang He, Jingtao Zhang, Jun Zhu, and Guihua Zeng.
\newblock Continuous-variable quantum teleportation in bosonic structured
  environments.
\newblock {\em Physical Review A}, 84(3):034305, 2011.

\bibitem{cuzminschi2021extractable}
Marina Cuzminschi, Alexei Zubarev, and Aurelian Isar.
\newblock Extractable quantum work from a two-mode gaussian state in a noisy
  channel.
\newblock {\em Scientific Reports}, 11(1):24286, 2021.

\bibitem{zubarev2019optimal}
Alexei Zubarev, Marina Cuzminschi, and Aurelian Isar.
\newblock Optimal fidelity of teleportation using two-mode gaussian states in a
  thermal bath as a resource.
\newblock {\em Rom. J. Phys}, 64:108, 2019.

\bibitem{marian2006continuous}
Paulina Marian and Tudor~A Marian.
\newblock Continuous-variable teleportation in the characteristic-function
  description.
\newblock {\em Physical Review A}, 74(4):042306, 2006.

\bibitem{schrodinger1935discussion}
Erwin Schr{\"o}dinger.
\newblock Discussion of probability relations between separated systems.
\newblock In {\em Mathematical Proceedings of the Cambridge Philosophical
  Society}, volume~31, pages 555--563. Cambridge University Press, 1935.

\bibitem{einstein1935can}
Albert Einstein, Boris Podolsky, and Nathan Rosen.
\newblock Can quantum-mechanical description of physical reality be considered
  complete?
\newblock {\em Physical review}, 47(10):777, 1935.

\bibitem{cuzminschi2021quantum}
Marina Cuzminschi and Aurelian Isar.
\newblock Quantum steering of two bosonic modes in the two-reservoir model.
\newblock {\em Romanian Reports in Physics}, 73(2), 2021.

\bibitem{wiseman2007steering}
Howard~M Wiseman, Steve~James Jones, and Andrew~C Doherty.
\newblock Steering, entanglement, nonlocality, and the einstein-podolsky-rosen
  paradox.
\newblock {\em Physical review letters}, 98(14):140402, 2007.

\bibitem{branciard2012one}
Cyril Branciard, Eric~G Cavalcanti, Stephen~P Walborn, Valerio Scarani, and
  Howard~M Wiseman.
\newblock One-sided device-independent quantum key distribution: Security,
  feasibility, and the connection with steering.
\newblock {\em Physical Review A}, 85(1):010301, 2012.

\bibitem{quintino2015inequivalence}
Marco~T{\'u}lio Quintino, Tam{\'a}s V{\'e}rtesi, Daniel Cavalcanti, Remigiusz
  Augusiak, Maciej Demianowicz, Antonio Ac{\'\i}n, and Nicolas Brunner.
\newblock Inequivalence of entanglement, steering, and bell nonlocality for
  general measurements.
\newblock {\em Physical Review A}, 92(3):032107, 2015.

\bibitem{jones2007entanglement}
Steve~James Jones, Howard~Mark Wiseman, and Andrew~C Doherty.
\newblock Entanglement, einstein-podolsky-rosen correlations, bell nonlocality,
  and steering.
\newblock {\em Physical Review A}, 76(5):052116, 2007.

\bibitem{piani2015necessary}
Marco Piani and John Watrous.
\newblock Necessary and sufficient quantum information characterization of
  einstein-podolsky-rosen steering.
\newblock {\em Physical review letters}, 114(6):060404, 2015.

\bibitem{giedke2002characterization}
G{\'e}za Giedke and J~Ignacio Cirac.
\newblock Characterization of gaussian operations and distillation of gaussian
  states.
\newblock {\em Physical Review A}, 66(3):032316, 2002.

\bibitem{fiuravsek2007gaussian}
Jarom{\'\i}r Fiur{\'a}{\v{s}}ek and Ladislav Mi{\v{s}}ta~Jr.
\newblock Gaussian localizable entanglement.
\newblock {\em Physical Review A}, 75(6):060302, 2007.

\bibitem{williamson1936algebraic}
John Williamson.
\newblock On the algebraic problem concerning the normal forms of linear
  dynamical systems.
\newblock {\em American journal of mathematics}, 58(1):141--163, 1936.

\bibitem{kogias2015quantification}
Ioannis Kogias, Antony~R Lee, Sammy Ragy, and Gerardo Adesso.
\newblock Quantification of gaussian quantum steering.
\newblock {\em Physical review letters}, 114(6):060403, 2015.

\bibitem{wallsquantum}
DF~Walls and GJ~Milburn.
\newblock Quantum optics, springer-verlag, berlin 1994.

\bibitem{xiang2015nonclassical}
Shao-Hua Xiang, Yu-Jing Zhao, Xi-Xiang Zhu, and Ke-Hui Song.
\newblock Nonclassical correlation dynamics in a system of mesoscopic josephson
  junction coupled to single-mode optical cavity.
\newblock {\em International Journal of Theoretical Physics}, 54:2881--2892,
  2015.

\end{thebibliography}
\end{document}